\documentstyle[proceedings]{crckapb}
\input epsf
\begin{opening}

\title{ ATMOSPHERE MODELS FOR VERY LOW MASS STARS, BROWN DWARFS AND EXOPLANETS}
\author{I. Baraffe}
\institute{Ecole Normale Sup\'erieure, CRAL, 69364 Lyon Cedex 07, France}
\author{F. Allard}
\institute{Dept. of Physics, Wichita State University,
Wichita, KS 67260-0032}
\end{opening}
\begin{document}

\section{Introduction}

Over the past decade considerable effort, both observational and
theoretical, has been directed towards a more accurate
determination of the stellar lower main sequence and of 
the sub-stellar domain covered by  Brown Dwarfs and Planets.
Astronomers have been looking for brown dwarfs for more than a decade, either
with standard astronomical technics
or with microlensing experiments. A breakthrough in the search for brown dwarfs
was very recently achieved with the discovery of the first cool brown dwarf
GL 229B (Nakajima et al. 1995). At the same epoch, the search for planets blossomed with the
discovery of a Jupiter - mass companion of the star 51 Pegasi (Mayor and Queloz 1995).
Now, the number of faint, cool stars and substellar objects is rising
rapidly. 

The most reliable way to identify the nature of observed objects
 is a direct comparison of observed and synthetic spectra, to  determine the effective temperature and metallicity. The mass  requires interior
models and is determined from consistent evolutionary models (cf. \S 2). 
Until recently, little about the atmospheres and spectral
characteristics of such cool objects was understood. 

\section{Characteristics of cool dwarf model atmospheres}

Very low mass stars (VLMS) or M-dwarfs are characterized by effective temperatures from  $\sim$ 5000 K down to 2000 K and surface gravities   $log \,  g\approx 3.5-5.5$, whereas 
 brown
dwarfs (BD) and extra-solar giant planets (EGP) can cover a much cooler temperature regime, down to some 100 K. Such low effective temperatures allow the presence of stable molecules (H$_2$, H$_2$O, TiO, VO, CH$_4$, NH$_3$,...), whose bands constitute
the main source of absorption along the characteristic frequency domain.
Such particular conditions are responsible for strong non-grey effects and  significant departure of the spectral energy distribution from a black body emission.
Tremendous progress has been made within the past years to derive
accurate non-grey atmosphere models by several groups over a wide range
of temperatures and metallicities (Plez et al. 1992; Saumon et al. 1994; Brett 1995
Allard and Hauschildt 1995,1997). A detailed description of the progress in the field is given in Allard et al. (1997). 

The current grids of atmosphere models now available do not however include  the condensation of molecules into grain, which should affect 
the atmosphere structure and the spectra of cool objects below T$_{eff} \sim$ 2500 K (Tsuji et al. 1996a). A recent breakthrough was achieved by Tsuji et al. (1996a)
who first included the grain formation
 and the grain opacities in non-grey atmosphere models for M-dwarfs and brown dwarfs. 
New generations of atmosphere
models taking into account molecular condensation are now essential 
for a correct description of  substellar objects and are now underway (e.g.  Burrows et al. 1996, Allard and Alexander 1997).
 
\begin{figure}
\epsfxsize=100mm
\epsfysize=80mm
\epsfbox{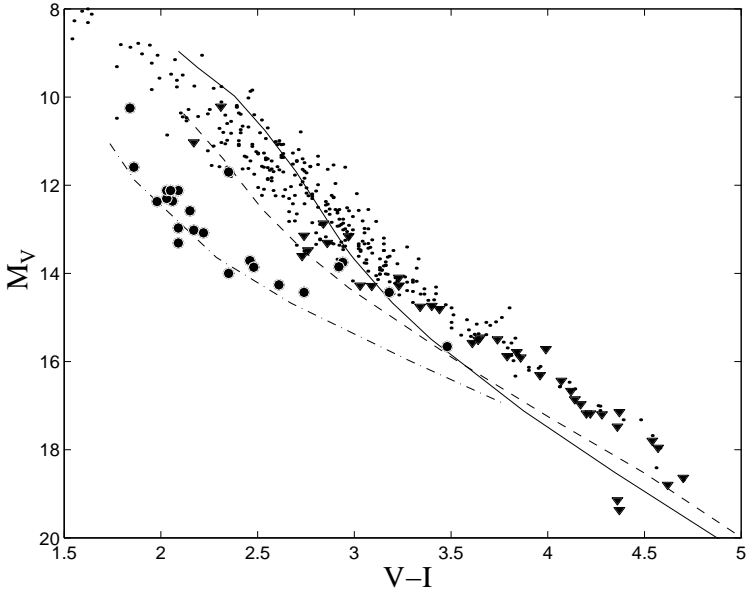}
\caption{Color - Magnitude diagrams for different metallicity: 
[M/H]=-1.5 (dash-dot), [M/H]=-0.5 (dash) and
[M/H]=0 (solid) from Baraffe et al. (1995). Subdwarf halo field stars from Monet et al. (1992) are indicated by
full circles, as well as disk stars of Monet et al. (1992) (full triangles) and Dahn et al. (1995) (dots) }
\end{figure}
 
Another difficulty inherent to cool dwarf atmopsheres is due to the presence of
convection in the optically thin layers. This particularity is due to the
molecular hydrogen recombination (H+H $\rightarrow$ H$_2$) which lowers the adiabatic gradient and favors the onset of convective instability.
Since radiative equilibrium is no longer satisfied in such atmospheres, 
the usual procedure based on $T(\tau)$
relationships to construct grey atmosphere models and to impose an outer 
boundary condition for the evolutionary models is basically incorrect (cf. Chabrier and
Baraffe 1997; 
Chabrier, this conference). 
An accurate surface
boundary condition based on {\it non-grey atmosphere models} is therefore required for evolutionary models. 

\section{Theory versus observation}

\subsection{M-dwarfs}

The main improvement  performed in the modelling of cool model atmospheres
is essentially due to more accurate molecular opacity treatment (e.g opacity sampling method, line by line treatment) and improved molecular data for the main absorbers as H$_2$O and TiO (Jorgensen 1994; Miller et al. 1994; etc...). 

The consistent coupling between non-grey atmosphere models and evolutionary
models and the use of synthetic spectra to derive the colors provided recently the clue to reproduce the disk and halo field stars
in a color-magnitude diagram (cf. Fig. 1). 
Same agreement is now reached with HST observations of Globular Cluster Main Sequences of low metallicity (Baraffe et al. 1997; see Chabrier 1997, this conference), based on the most recent generation of
atmosphere models (Allard et al. 1997). These successful results suggest that the present  stellar models are now sufficiently accurate to derive reliable mass-luminosity relationships and thus mass functions, although for solar metallicity a problem remains for the very bottom of the Main Sequence ( $\le
0.1 M_\odot$).

For solar metallicity, an empirical mass-luminosity relationship has been derived by Henry and McCarthy (1993, HMC93) based on observations of binary systems. Improvement in the molecular linelists of TiO and H$_2$O now leads to a good agreement of the models with the observed relationship in the optical (V-band) and in the near IR
(K-band). Figure 2 shows such agreement for stellar models based on recent
atmosphere models of Allard and Hauschildt (1997) and Brett-Plez (Brett 1995;
Chabrier et al. 1996). Note that the change of slope in the HMC93 fit at
M$_V \sim 11$ is exagerated and stems from a linear interpolation between remote objects (see Fig. 2 of HMC93).

\begin{figure}
\epsfxsize=100mm
\epsfysize=80mm
\epsfbox{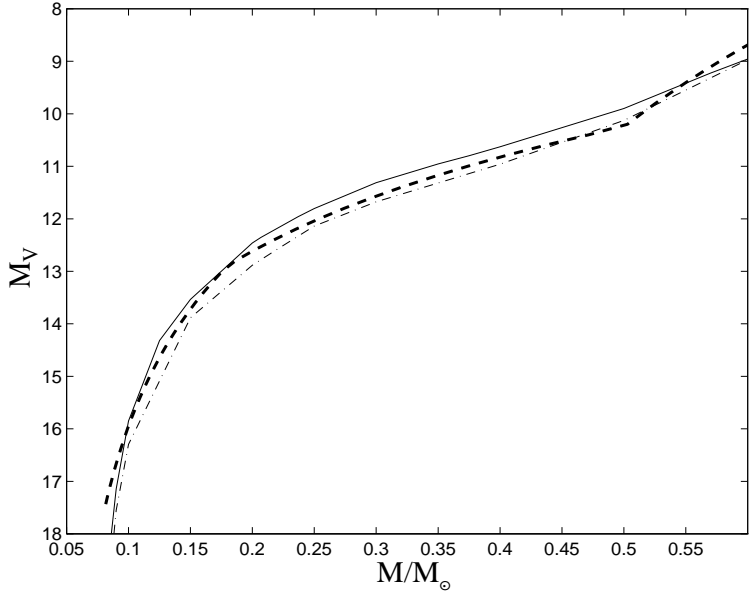}
\caption{Mass - Magnitude relationship in the V-band. The thick dashed line
corresponds to the empirical relationship derived by Henry and McCarthy (1993).
Evolutionary  models (Chabrier et al. 1996) are based on the Allard and Hauschildt
(1997 , solid line) and the Brett-Plez (dash-dot) non-grey atmosphere models.  }
\end{figure}

The comparison of synthetic spectra with observed spectral distribution of M-dwarfs now reaches a good agreement in the IR (cf. Leggett et al. 1996; Allard et al. 1997) for atmosphere models including the new water linelist of Miller et al. (1994). Disagreement still remains in the optical colors (V-I, R-I)
and for the latest-type M-dwarfs (T$_{eff} \le 3500$ K) where the models
systematically overestimate the water band strengh. Such discripancies may be due either to remaining shortcomings in the TiO (optical) and H$_2$O (IR) line lists or to the effect of grain formation, as discussed below.

\subsection{Substellar objects}

As the effective temperature decreases below $\sim$ 2000 K, the IR spectrum
is dominated mainly by H$_2$O, CH$_4$ and NH$_3$. The presence of methane is a clear signature of the substellar nature of an object.  
The most stringent
test for extremely cool models was provided recently by the brown dwarf Gl 229B (Nakajima et al. 1996). The synthetic spectra calculated by different groups for an effective temperature of $\sim 1000 K$ yield to a remarkable agreement with the observed spectrum, reproducing the main molecular absorption features
(Allard et al. 1996, Tsuji et al. 1996b, Marley et al. 1996). 

The predicted absolute fluxes of BD or EGP (Allard et al. 1997)
are illustrated in Fig. 3 and
compared to the sensitivity of ground and space-based observing platforms, 
as given by Saumon et al. (1996).
The predictions are shown for different effective temperatures and
compared to a black body emission at the same T$_{eff}$. 
Strong departure from the
black body emission illustrates the necessity to derive synthetic spectra
for this type of objects. As shown in Fig. 3, the flux peaks in the 4.5 - 5
$\mu m$ window, which seems to be the best region for the search of
cool BD or EGP. 
 
\begin{figure}
\epsfxsize 100mm
\epsfysize 80mm
\epsfbox{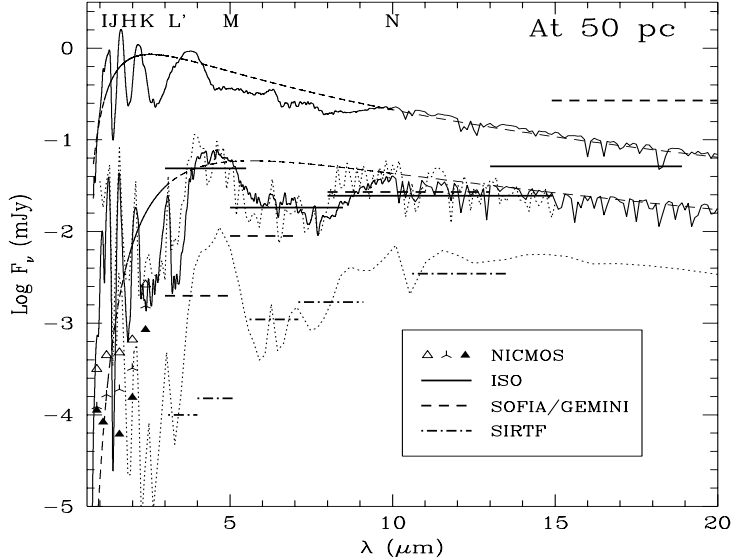}
\caption{Predicted absolute fluxes of BD or EGP at 50 pc compared to the sensitivity of ground and space-based platforms. Models of both
Allard et al. (1996) (solid) and Marley et al. (1996) (dotted) are shown
which simulate (i) a brown dwarf near the hydrogen burning limit (topmost
spectrum $T_{eff} = 2000K$), (ii) an evolved brown dwarf similar to GL 229B (central spectra: $T_{eff} = 900K$ and 960K), and (iii) an extremely cool object (lowermost spectrum: $T_{eff} = 500K$). Dashed curves give the corresponding black-body. }
\end{figure}

Despite the success of modelling GL 229B, the remaining challenge for atmosphere
modellers in the low temperature regime is the grain formation. 
Although the effects of grain absorption appear
more subtil in currently available IR spectra of Gl229B (cf. Tsuji et al. 1996b;  Allard et al. 1996, Marley et al. 1996),
the inclusion of this effect
 seems necessary to explain the IR spectra of very late-type M-dwarfs. 
As shown by Tsuji et al. (1996a) and illustrated in Figure 4, kindly provided by T. Tsuji, dust opacities tend to reduce the molecular
absorption feature and could explain the overestimation of 
water absorption found in  dust-free model atmospheres.

\begin{figure}
\epsfxsize 100mm
\epsfysize 90mm
\epsfbox{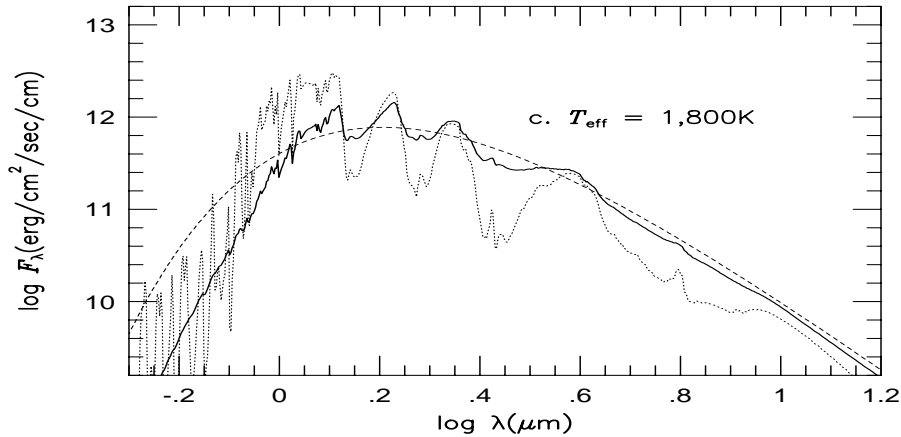}
\caption{ Predicted spectral energy distribution for dusty (solid line) and
dust-free (dotted line) models with T$_{eff}$ = 1800K (from Tsuji et al. 1996a). The black body
curve corresponds to the dashed line. }
\end{figure}

\section{Conclusion}

The last few years have shown impressive improvement in the modelization of
cool model atmospheres. The main efforts in both theoretical and observational
directions lead to a much better understanding of the bottom of the
Main Sequence and the substellar regime. Remarkable agreements between theory
and observations are now reached on different fronts: color - magnitude diagram, mass - luminosity relationships, spectrum. The success of the theory is most encouraging and shows that it has finally come to maturity.

\end{document}